\documentclass[aps,prl,reprint,groupedaddress]{revtex4-1}
\usepackage{graphicx} 
\usepackage{dcolumn} 
\usepackage{bm}
\usepackage{hyperref}

\begin{document}

\title{Systematics on production of superheavy nuclei $Z = 119-122$ in fusion-evaporation reactions}
\author{Fei Niu }
\affiliation{School of Physics and Optoelectronic Technology, South China University of Technology, Guangzhou 510640, China }

\author{Peng-Hui Chen }
\affiliation{College of Electrical Energy and Power Engineering, Yangzhou University, Yangzhou 225000, China }

\author{Zhao-Qing Feng }
\email{Corresponding author: fengzhq@scut.edu.cn}
\affiliation{School of Physics and Optoelectronic Technology, South China University of Technology, Guangzhou 510640, China }

\date{\today}

\begin{abstract}
The fusion dynamics on the formation of superheavy nuclei is investigated thoroughly within the dinuclear system model. The Monte Carlo approach is implemented into the nucleon transfer process for including all possible orientations, at which the dinuclear system is assumed to be formed at the touching configuration of dinuclear fragments. The production cross sections of superheavy nuclei Cn, Fl, Lv, Ts and Og are calculated and compared with the available data from Dubna. The evaporation residue excitation functions in the channels of pure neutrons and charged particles are analyzed systematically. The combinations with $^{44}$Sc, $^{48,50}$Ti, $^{49,51}$V, $^{52,54}$Cr, $^{58,62}$Fe and $^{62,64}$Ni bombarding the actinide nuclides $^{238}$U, $^{244}$Pu, $^{248}$Cm, $^{247,249}$Bk, $^{249,251}$Cf, $^{252}$Es and $^{243}$Am are calculated for producing the superheavy elements with Z=119-122. It is found that the production cross sections sensitively depend on the neutron richness of reaction system. The structure of evaporation residue excitation function is related to the neutron separation energy and fission barrier of compound nucleus.

\begin{description}
\item[PACS number(s)]
25.70.Gh, 25.70.Jj, 27.90.+b
\end{description}
\end{abstract}

\maketitle

\section{Introduction}

Over the past decades, the synthesis of superheavy nuclei (SHN) attracted much attention and has been obtained progress in experiments via massive fusion reactions. The seventh period in the periodic table was filled with the superheavy element tennessine (Ts) by using the Dubna Gas-Filled Recoil Separator (DGFRS) at the Flerov Laboratory of Nuclear Reactions in Dubna, Russia \cite{Og10}. The existence of superheavy elements (SHE) was predicted in the late 1960s by the macroscopic-microscopic theory of the atomic nucleus \cite{So66}. The synthesis of SHN is associated with testing the shell model beyond the doubly magic nucleus $^{208}$Pb, hunting for the 'island of stability', exploring the limit of the mass of atomic nucleus, providing a strong Coulomb field such as quantum electrodynamics (QED) in the super-strong electric field etc. The existence of the superheavy nucleus (SHN) ($Z\geq106$) is due to strong binding shell effect against the Coulomb repulsion. Therefore, the position of shell closure is particularly significant for the properties of SHN, i.e., half-lives of $\alpha$ decay chain and spontaneous fission, formation probability etc, which is predicted at Z=114 and N=184 by theoretical models \cite{So07,Cw05}. Attempts to synthesize elements beyond Og (Z=118) were performed via different systems, e.g., $^{64}$Ni+$^{238}$U \cite{Ho08}, $^{58}$Fe+$^{244}$Pu \cite{Og09}, $^{54}$Cr+$^{248}$Cm \cite{Ho12,Ho16} and $^{50}$Ti+$^{249}$Cf \cite{Kh13}. The mass and angle distributions of fission fragments were measured \cite{Al20}. Systematical analysis of different reactions needs to be performed for preferentially producing new SHN in experiments.

The synthesis of SHN goes back 40 years ago with the multi-nucleon transfer reactions in collisions of two actinide nuclei \cite{Hu77,Sc78}. However, the yields of the heavy fragments in strongly damped collisions were found to decrease very rapidly with increasing the atomic number and to be impossible for producing SHN because of the very low cross section. Combinations with a doubly magic nucleus or nearly magic nucleus are usually chosen owing to the larger reaction $Q$ values. Reactions with $^{208}$Pb or $^{209}$Bi based targets were firstly proposed by Oganessian et al. \cite{Og75}. The SHEs from Bh to Cn were synthesized in the cold fusion reactions at GSI (Darmstadt, Germany) with the heavy-ion accelerator UNILAC and the SHIP separator \cite{Ho00,Mu15}. Experiments on the synthesis of element Nh (Z=113) in the $^{70}$Zn+$^{209}$Bi reaction have been performed successfully at RIKEN (Tokyo, Japan) \cite{Mo04}. However, it is very difficulty to create the superheavy isotopes beyond Nh in the cold fusion reactions because of very low cross section ($\sigma<$0.1 pb). The superheavy elements from Fl (Z=114) to Og (Z=118) have been synthesized at the Flerov Laboratory of Nuclear Reactions (FLNR) in Dubna (Russia) with the double magic nuclide $^{48}$Ca bombarding actinide nuclei \cite{Og99,Og06,Og15}, in which more neutron-rich SHN were produced and identified by the subsequent $\alpha$-decay chain. The decay properties of $^{271}$Ds in the cold fusion reaction of $^{64}$Ni+$^{208}$Pb$\rightarrow ^{271}$Ds+n were identified by a gas-filled recoil separator at Institute of Modern Physics (IMP) in Lanzhou \cite{Zh12}. With constructing the new facilities in the world such as RIBF (RIKEN, Japan), SPIRAL2 (GANIL in Caen, France), FRIB (MSU, USA), HIAF (IMP, China), it would be possible to create SHNs on the 'island of stability' by using the very neutron-rich radioactive beams in the near future.

The formation dynamics of SHN in the massive fusion and multinucleon transfer reactions is complicated, which is associated with the coupling of several degrees of freedom, such the radial elongation, mass or charge asymmetry, shape configuration, relative motion energy etc. Several macroscopic models are developed for describing the fusion hindrance in massive systems, e.g., the macroscopic dynamical model \cite{Bj82}, fusion-by-diffusion (FBD) model \cite{Sw05}, dynamical models based on Langevin-type equations \cite{Za07,Li11}, dinuclear system (DNS) model \cite{Ad97,Fe06,Fe11} etc. Recently, the time-dependent Hartree-Fock (TDHF) method was also applied to investigate the quasifission and fusion-fission dynamics in the reactions of $^{48}$Ca+$^{239,244}$Pu \cite{Gu18}. Modifications of macroscopic models are still needed for self-consistently and reasonably explaining the fusion dynamics in massive systems. The production cross sections of SHEs Z=119 and 120 were estimated within the multidimensional Langevin-type equations \cite{Za15} and DNS model \cite{Na09,Fe09,Ga11,Wa12,Li18,Ad20} for different reaction systems. Systematic study on the SHN production beyond oganesson (Z=118) is needed for predicting the optimal projectile-target combinations and reaction mechanism.

In this work, the stochastic diffusion in the nucleon transfer process is taken into the DNS model via the Monte Carlo procedure. The systematic analysis on the production of new superheavy elements is performed. The paper is organized as follows. In Sec. II, a brief description of the DNS model is presented. The comparison with available data and predictions of new elements Z=119-122 are discussed in Sec. III. A summary is given in IV.

\section{Model description}

The DNS model has been applied to the the quasi-fission and fusion dynamics, multinucleon transfer reactions and deep inelastic collisions, in which the dissipation of relative motion and rotation of colliding system into the internal degrees of freedom is assumed at the touching configuration. The DNS system evolves along two main degrees of freedom to form a compound nucleus, namely, the radial motion via the decay of DNS and the nucleon transfer via the mass asymmetry $\eta=(A_{1}-A_{2})/(A_{1}+A_{2})$ \cite{Ad20b,Fe09b}. In accordance with the temporal sequence, the system undergoes the capture by overcoming the Coulomb barrier, the competition of quasi-fission and complete fusion by cascade nucleon transfer, and the formation of cold residue nuclide by evaporating $\gamma$-rays, neutrons, light charged particles and binary fission.
The production cross section of the superheavy residue is estimated by the sum of partial wave with the angular momentum $J$ at incident center of mass energy $E_{c.m.}$ as,
\begin{eqnarray}
\sigma_{ER}(E_{c.m.}) = \frac{\pi\hbar^{2}}{2\mu E_{c.m.}}\sum^{J_{max}}_{J=0}(2J+1)T(E_{c.m.},J) P_{CN}(E_{c.m.},J)W_{sur}(E_{c.m.},J)
\end{eqnarray}
Here, $T(E_{c.m.},J)$ is the penetration probability and given by a Gaussian-type barrier distribution. The fusion probability $P_{CN}$ is described by the DNS model and taking into account the competition of the quasi-fission and fission of the heavy fragment. The survival probability $W_{sur}$ is calculated with the statistical approach \cite{Ch17,Ch18}.

In the DNS model, the time evolution of the distribution probability $P(Z_{1},N_{1},E_{1}, t)$ for the fragment 1 with proton number $Z_{1}$, neutron number $N_{1}$ and excitation energy $E_{1}$ is described by the following master equations as
\begin{eqnarray}
&& \frac{d P(Z_{1},N_{1},E_{1}, t)}{dt}           \nonumber\\
  &=& \sum_{Z^{'}_{1}}W_{Z_{1},N_{1};Z^{'}_{1},N_{1}}(t)[d_{Z_{1},N_{1}}P(Z^{'}_{1},N_{1},E^{'}_{1},t)    \nonumber\\
 && - d_{Z^{'}_{1},N_{1}}P(Z_{1},N_{1},E_{1},t)] + \sum_{N^{'}_{1}}W_{Z_{1},N_{1};Z_{1},N^{'}_{1}}(t)     \nonumber\\
 && \times [d_{Z_{1},N_{1}}P(Z_{1},N^{'}_{1},E^{'}_{1},t)  - d_{Z_{1},N^{'}_{1}}P(Z_{1},N_{1},E_{1}, t)]    \nonumber\\
 && -[\Lambda^{qf}_{Z_{1},N_{1},E_{1},t}(\Theta)+\Lambda^{fis}_{Z_{1},N_{1},E_{1},t}(\Theta)]P(Z_{1},N_{1},E_{1},t).   \nonumber\\
 &&
\end{eqnarray}
Here the $W_{Z_{1},N_{1};Z^{'}_{1},N_{1}}$($W_{Z_{1},N_{1};Z_{1},N^{'}_{1}}$) is the mean transition probability from the channel($Z_{1},N_{1},E_{1}$) to ($Z^{'}_{1},N_{1},E^{'}_{1}$), [or ($Z_{1},N_{1},E_{1}$) to ($Z_{1},N^{'}_{1},E^{'}_{1}$)], and $d_{Z_{1},Z_{1}}$ denotes the microscopic dimension corresponding to the macroscopic state ($Z_{1},N_{1},E_{1}$).The sum is taken over all possible proton and neutron numbers that fragment $Z^{'}_{1}$, $N^{'}_{1}$ may take, but only one nucleon transfer is considered in the model with the relations $Z^{'}_{1}=Z_{1}\pm1$ and $N^{'}_{1}=N_{1}\pm1$. The excitation energy $E_{1}$ is determined by the dissipation energy from the relative motion and the potential energy surface of the DNS. The quasi-fission rate $\Lambda^{qf}$ and fission rate $\Lambda^{fis}$ are given by the one-dimensional Kramers formula \cite{Ad03}. The motion of nucleons in the interacting potential is governed by the single-particle Hamiltonian \cite{Fe06}, which is influenced by the local excitation energy of DNS.

The potential energy surface (PES) of the DNS is given by
\begin{eqnarray}
U(\{\alpha\}) =  B(Z_{1},N_{1})+B(Z_{2},N_{2}) - \left[B(Z,N)+V^{CN}_{rot}(J)\right]   +  V(\{\alpha\}).
\end{eqnarray}
The DNS fragments satisfy the relation of $ Z_{1}+Z_{2}=Z $ and  $ N_{1}+N_{2}=N$ with the $Z$ and $N$ being the proton and neutron numbers of composite system, respectively. The symbol ${\alpha}$ denotes the quantities of $Z_{1}$, $N_{1}$, $Z_{2}$, $N_{2}$, $J$, $R$, $\beta_{1}$, $\beta_{2}$, $\theta_{1}$, $\theta_{2}$. The $B(Z_{i},N_{i}) (i=1, 2)$ and $B(Z,N)$ are the negative binding energies of the fragment $(Z_{i},N_{i})$ and the compound nucleus $(Z,N)$, respectively. The $V^{CN}_{rot}$ is the rotation energy of the compound nucleus. The $\beta_{i}$ represent the quadrupole deformations of the two fragments and are taken the ground-state values. The $\theta_{i}$ denote the polar angles between the collision orientations and the symmetry axes of the deformed nuclei. The collision direction is sampled via the Monte Carlo method with the $\theta_{i}=\frac{\pi}{2}\xi_{i}$ with $\xi_{i}$ being the random number. It should be noticed that the angles $\theta_{i} (i=1, 2)$ of binary DNS fragments are stochastically sampled and different in the nucleon transfer process. The interaction potential between fragments $(Z_{1},N_{1})$ and $(Z_{2},N_{2})$ includes the nuclear, Coulomb, and centrifugal parts as
\begin{eqnarray}
V(\{\alpha\})=V_{N}(\{\alpha\})+\frac{J(J+1)\hbar^{2}}{2\mu R^{2}}+V_{C}(\{\alpha\})
\end{eqnarray}
$\mu$ is the reduced mass of two DNS fragments. The nuclear potential is calculated by using the double-folding method based on the Skyrme interaction force without considering the momentum and the spin dependence \cite{Ad39}. The Coulomb potential is obtained by the Wong's formula \cite{Wo40}. In the calculation, the distance $R$ between the centers of the two fragments is chosen as the minimal position of interaction pocket. The minimal path in the valley of PES is named as the driving potential and dependent on the mass asymmetry. Shown in Fig. 1 is the driving potentials for the tip-tip, waist-waist collisions and average value of random orientations in the reaction of $^{48}$Ca + $^{238}$U. The tip-tip orientation means that the interaction potential is the minimal value with varying the polar angle $\theta$ corresponding to the inner fusion barrier of $B_{fus}= 9.9$ MeV, and the waist-waist case being the maximal potential for the DNS fragments with $B_{fus}= 4.8$ MeV. In all orientations, the driving potential manifests the symmetric structure. It is obvious that the driving potential with the random collisions is basically close to the average values of the tip-tip and waist-waist collisions. The inner fusion barrier is related to the collision orientation and the waist-waist collisions undergoes a high barrier to form a compound nucleus. However, the bump structure towards the decrease of mass asymmetry hinders the quasifission process.

\begin{figure*}
\includegraphics[width=16 cm]{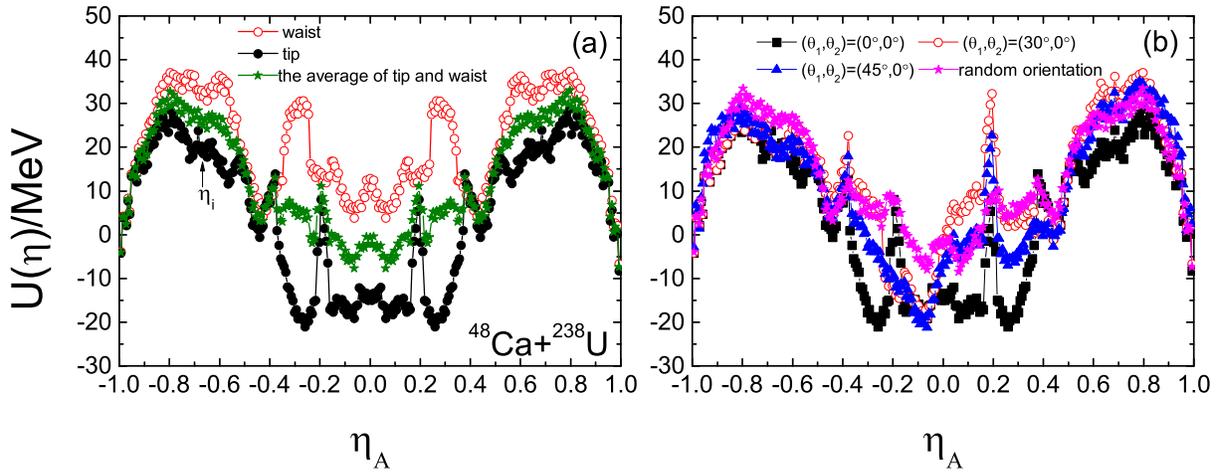}
\caption{The driving potentials in the reaction of $^{48}$Ca+$^{238}$U with the waist-waist, tip-tip, random collisions and fixed angles, respectively. }
\end{figure*}

In order to form a compound nucleus (CN), the DNS must have enough local excitation to overcome the internal fusion barrier. The formation probability of the compound nucleus at Coulomb barrier $B$ and angular momentum $J$ is given by the summation to the B.G. (Businaro-Gallone) point as
\begin{eqnarray}
P_{CN}(E_{c.m.},J,B) =  \frac{1}{N_{t}} \sum_{i=1}^{N_{t}} \sum^{Z_{B.G.}}_{Z_{1}=1}\sum^{N_{B.G.}}_{N_{1}=1}    P(Z_{1},N_{1},E_{1}(\theta_{1},\theta_{2}),\tau_{int}(\theta_{1},\theta_{2}))   \sin(\theta_{1}) \sin(\theta_{2})
\end{eqnarray}
Here the interaction time $\tau_{int}(E_{c.m.},J,B)$ is obtained by using the deflection function method \cite{Li41}. The excitation energy $E_{1}$ of DNS fragment ($Z_{1}, N_{1}$) is related to the collision orientation $\theta_{1}$ and $\theta_{2}$. The $N_{t}$ is the total event for Monte Carlo integration with $\theta_{i}=\frac{\pi}{2}\xi$. The fusion probability is calculated with a Gaussian-type distribution $f(B)$ as
\begin{eqnarray}
P_{CN}(E_{c.m.},J) = \int f(B) P_{CN}(E_{c.m.},J,B)dB.
\end{eqnarray}
The collision orientation influences the PES of DNS because of different interaction potential with the stochastic angle. Consequently, the formation probability of CN is related to the orientation of two DNS fragments. Shown in Fig. 2 is the dependence of the fusion probability on the excitation energy at different orientation in nucleon transfer process in the reaction of $^{48}$Ca+$^{238}$U. It is obvious that the fusion probability increases with the excitation energy of compound nucleus. The waist-waist case leads to the high fusion probability owing to the lower inner fusion barrier and a small peak towards the symmetric diffusion (quasifission path). The statistical error is included for the random collisions, in which the fusion probability lies between the fixed orientations. Usually, the tip-tip orientation is chosen in the calculation of fusion probability, which is considered to be probable nucleon transfer \cite{Fe06}.

\begin{figure*}
\includegraphics[width=14 cm]{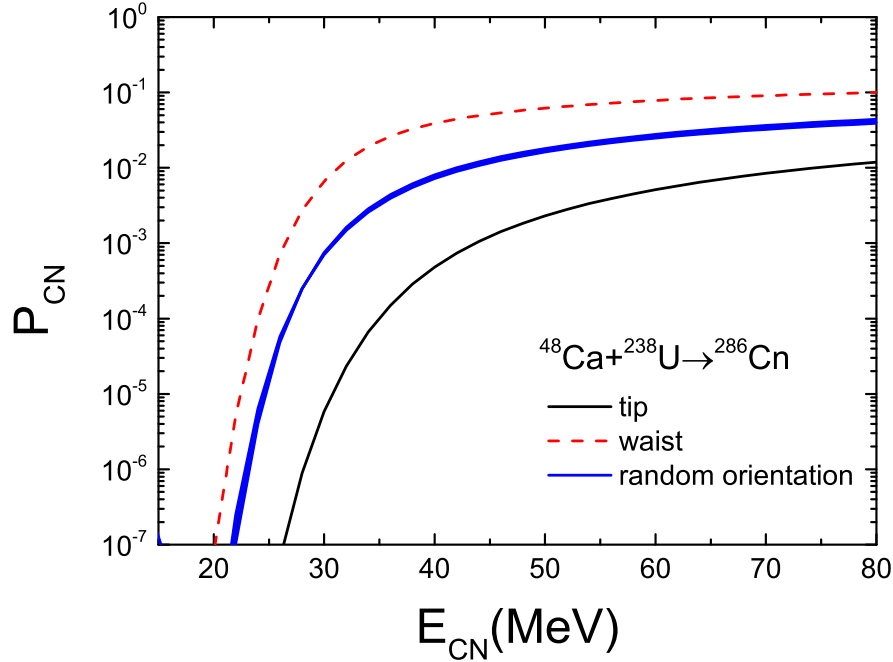}
\caption{ A comparison of the fusion probability a function of excitation energy in the reaction of $^{48}$Ca+$^{238}$U at different collision orientations. }
\end{figure*}

Once the compound nucleus is formed by the cascade nucleon transfer, the de-excitation undergoes by emitting $\gamma$-rays, particles (n, p, d, $\alpha$ etc) and binary fission. The survival probability of heavy nuclei after evaporating particle is very important for assessment of cross section, which is usually calculated with the statistical process. The probability in the channel of evaporating the $x$-th neutron, the $y$-th proton and the $z$th-$\alpha$ is expressed as \cite{Ch17}
\begin{eqnarray}
&& W_{sur}(E^{*}_{CN},x,y,z,J)=P(E^{*}_{CN},x,y,z,J)     \nonumber  \\
&&   \times\prod^{x}_{i=1}\frac{\Gamma_{n}(E^{*}_{i},J)}{\Gamma_{tot}(E^{*}_{i},J)}\prod^{y}_{j=1}\frac{\Gamma_{p}(E^{*}_{j},J)}{\Gamma_{tot}(E^{*}_{j},J)}
\prod^{z}_{k=1}\frac{\Gamma_{\alpha}(E^{*}_{k},J)}{\Gamma_{tot}(E^{*}_{k},J)}.
\end{eqnarray}
The $E^{*}_{CN}$, $J$ and $\Gamma_{tot}$ are the excitation energy, the spin of compound nucleus and the sum of partial widths of particle evaporation, respectively. The excitation energy $E^{*}_{s}$ before evaporating the $s$-th particles is given by
\begin{eqnarray}
E^{*}_{s+1}=E^{*}_{s}-B^{n}_{i}-B^{p}_{j}-B^{\alpha}_{k}-2T_{s}
\end{eqnarray}
with the initial value $E^{*}_{1}=E^{*}_{CN}$ and $s=i+j+k$. The nuclear temperature $T_{i}$ is given by $E^{*}_{i}=aT^{2}_{i}-T_{i}$ with $a$ being the level density parameter. The widths of neutron evaporation and fission are calculated by using the Weisskopf evaporation theory. The fission barrier is evaluated from the macroscopic liquid drop model and the shell correction energy, and which is given as
\begin{equation}
B_{f}(E^{*},J)=B^{LD}_{f}+E_{shell}\exp(-E^{*}/E_{D}),
\end{equation}
where the macroscopic part $B^{LD}_{f}$ is calculated by liquid-drop model. The microcosmic shell correction energy is calculated with the Strutinsky method and obtained from Ref. \cite{Mo95}. The damping energy $E_{D}$ is associated with the level density and mass number of compound nucleus, in which the shell correction and excitation energy dependence are taken into account in the calculation \cite{Ch17}.

\section{Results and discussion}

The production rate of SHNs in massive fusion reactions is very low owing to the fusion hindrance in heavy systems, which enables the quasifission process in the binary collisions. The evaporation residue (ER) excitation functions in different channels are favorable for experimental measurements with optimal projectile-target combination and suitable beam energy. The reaction dynamics in competition of the quasifission and fusion-fission reactions, level density, separation energy of evaporated particle and fission barrier of compound nucleus influence the ER cross sections. As a test of the model, shown in Fig. 3 is the ER excitation functions in the reaction of $^{48}$Ca+$^{238}$U and compared with the data from Dubna \cite{Og43}. The ER cross sections for producing SHN strongly depend on the orientations of both DNS fragments in the nucleon transfer process. The tip-tip collisions have the lower cross sections in comparison with the waist-waist orientations and roughly lead to the two order reduction because of the higher inner fusion barrier for merging the compound nucleus. The cross sections with the stochastic selection of collision angle of two DNS fragments lie between the tip-tip and waist-waist orientation. The results with the tip-tip collisions are consistent with the available data and are chosen in the calculation. It is obvious that the maximal yield of SHN $^{283}$Cn production with 3 pb is positioned in the 3n channel via the tip-tip collisions at the excitation energy of 35 MeV. The SHN is still far away from the neutron shell closure N=184. New reaction mechanism is expected for creating the neutron-rich SHN. The pure neutron channels are dominant decay modes for surviving SHN. The cross sections with mixed channels of proton and $\alpha$ are reduced to be two order magnitude with the 3 particle channels, i.e., p2n, $\alpha$2n. The charged particle channels are of significance on the production of proton-rich actinide nuclides close to the drip line in the fusion-evaporation reactions and the multinucleon transfer dynamics \cite{Ch20}.

\begin{figure*}
\includegraphics[width=14 cm]{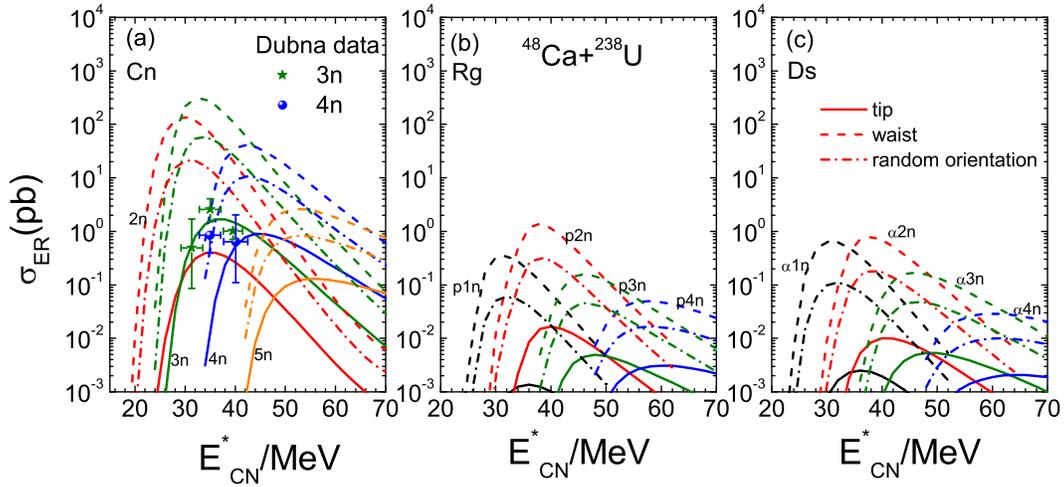}
\caption{ Calculated ER cross sections with different channels of (2-5)n, 1p(2-4)n and $1\alpha$(1-3)n and compared with the experimental data in the reaction of $^{48}$Ca+$^{238}$U \cite{Og43}.}
\end{figure*}

\begin{figure*}
\includegraphics[width=14 cm]{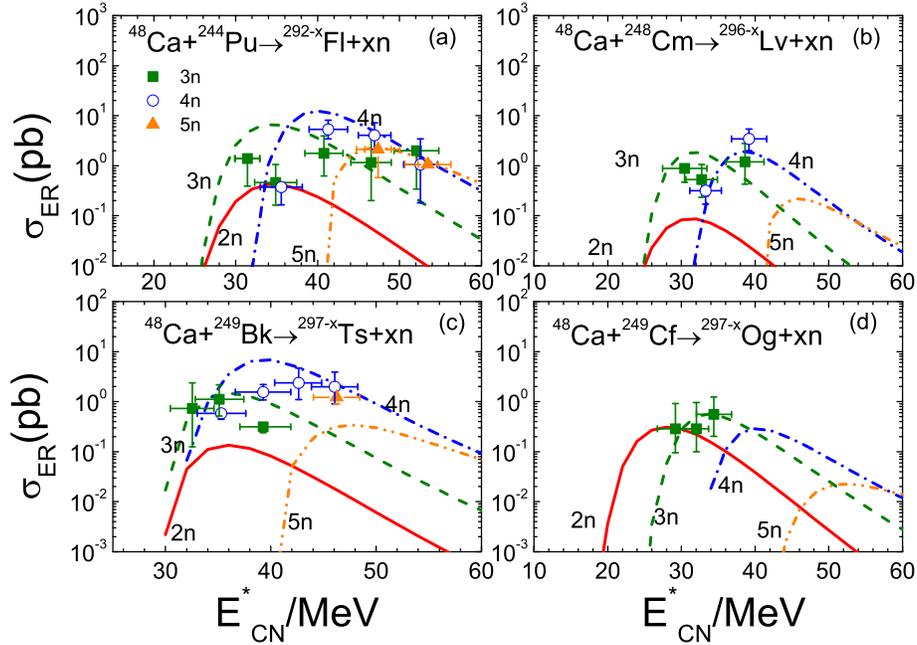}
\caption{ ER excitation functions with pure neutron channels and compared with the experimental data in the reactions of $^{48}$Ca+$^{244}$Pu, $^{248}$Cm, $^{249}$Bk, $^{249}$Cf \cite{Og43,Og44}. }
\end{figure*}

The superheavy elements from Fl (Z=114) to Og (Z=118) have been successfully synthesized with $^{48}$Ca induced reactions on actinide targets. It manifests the strong shell effect in the production and the decay chains. Shown in Fig. 4 is the ER excitation functions for producing the superheavy elements 114-118 with $^{244}$Pu, $^{248}$Cm, $^{249}$Bk and $^{249}$Cf as the targets. Different evaporation channels are distinguished by colored lines, and compared with the experimental data \cite{Og43,Og44}. It is obvious that the 3n and 4n channels are dominant decay modes for SHN production. The 2n channel is pronounced with increasing the charge number of SHN. However, the cross sections in the 4n and 5n channels decrease with becoming heavier SHN. Different with the cold fusion reactions, the maximal cross section in the $^{48}$Ca induced reactions weakly depends on the mass of ER nucleus, e.g., the value of 8 pb for $^{288}$Fl and 0.6 pb for $^{294}$Og. The difficulty is the construction of target material for experiments to synthesize new SHN. The hot fusion reactions also provide a possible way for creating the new elements at the eighth period with projectiles $^{45}$Sc, $^{50}$Ti, $^{54}$Cr, $^{58}$Fe, $^{64}$Ni etc. Besides the reaction dynamics in the formation of SHN in the hot fusion reactions, the nuclear structure effects are of importance in the evaluation of production cross section, i.e., shell effect, neutron separation energy, odd-even effect, microscopic state of compound nucleus (isomeric state) etc.

\begin{figure*}
\includegraphics[width=14 cm]{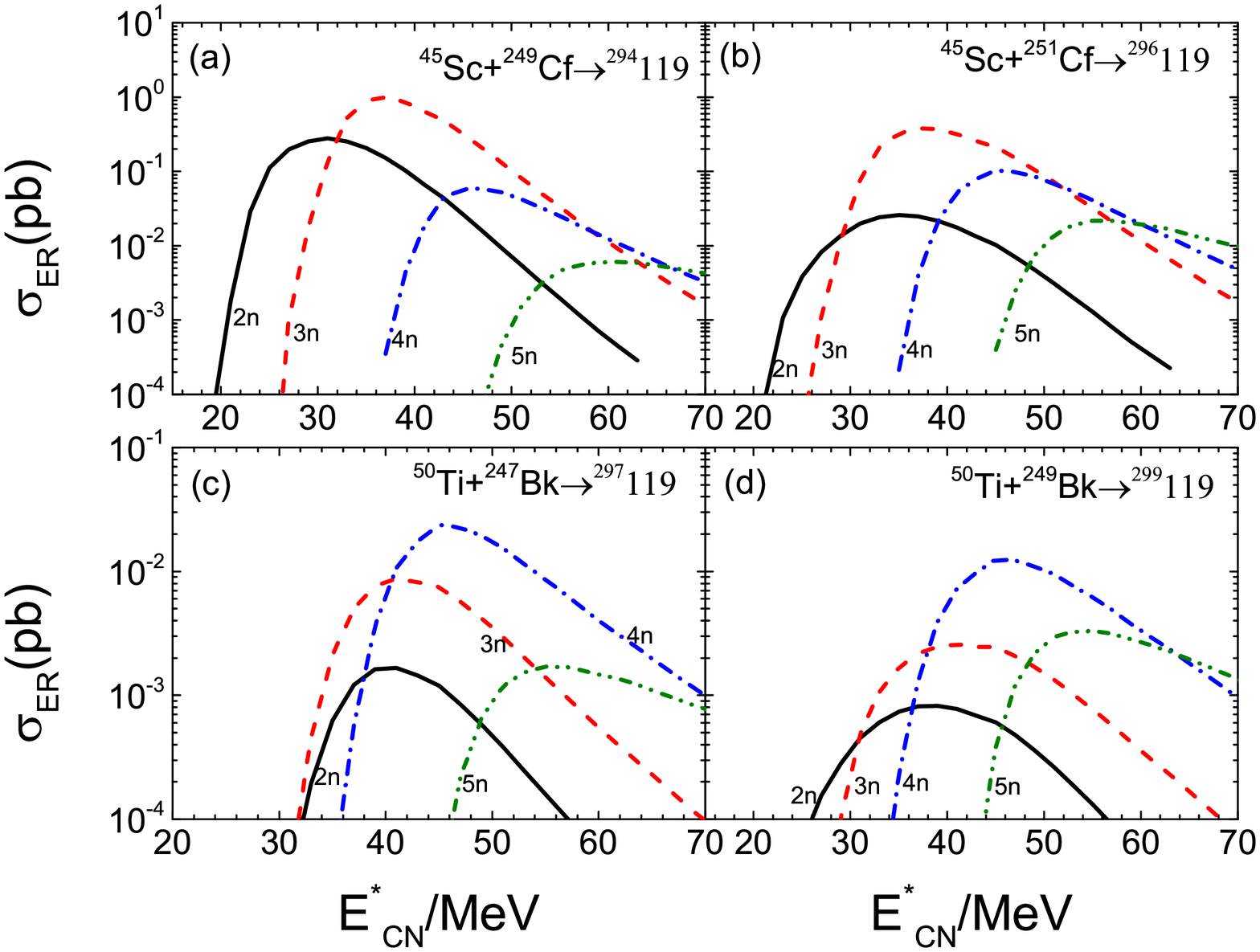}
\caption{ The evaporation residue cross sections with channels of (2-5)n of element Z=119 in collisions by different reactions.}
\end{figure*}

\begin{table}[!htb]
	\centering
	\caption{Optimal evaporation residual cross sections via different reactions leading to the formation of SHE Z=119. }
	\label{tab1}
		\begin{tabular}{ccccccc}
	Reaction systems  &$\sigma_{ER}(pb)$    &$E_{CN}^{\ast}$(MeV)       &References  \\
        \hline
            $^{249}$Bk($^{50}$Ti,3n)$^{296}119$    &0.04    &41           & \cite{Za08}   \\
            $^{249}$Bk($^{50}$Ti,4n)$^{295}119$    &0.06    &44          & \cite{Za08}  \\
            $^{254}$Es($^{48}$Ca,3n)$^{299}119$  &0.3       &35          &\cite{Za12}   \\
	    	$^{252}$Es($^{48}$Ca,4n)$^{296}$119  &0.2       &43          &\cite{Si46}   \\
			$^{254}$Es($^{48}$Ca,4n)$^{298}119$  &0.015   &41           &\cite{Si46}    \\
            $^{249}$Bk($^{50}$Ti,4n)$^{295}119$   &0.03     &36          &\cite{Si46}    \\
            $^{249}$Bk($^{50}$Ti,3n)$^{296}119$  &0.035   &27          &\cite{Wa12}  \\
            $^{249}$Bk($^{50}$Ti,4n)$^{295}119$  &0.11    &39            &\cite{Wa12}  \\
            $^{249}$Bk($^{50}$Ti,4n)$^{295}119$  &0.57   &41            &\cite{Liu11}  \\
            $^{252}$Es($^{44}$Ca,3n)$^{293}119$ &4.32  &35            &\cite{Li18}  \\
            \cline{1-4}
            $^{251}$Cf($^{45}$Sc,3n) $^{293}119$ &0.38  &37     & This work\\
            $^{249}$Cf($^{45}$Sc,3n)$^{291}119$ &0.99   &37      \\
            $^{247}$Bk($^{50}$Ti,4n)$^{293}119$ &0.024  &45    \\
            $^{249}$Bk($^{50}$Ti,4n)$^{295}119$ &0.013   &45

            \end{tabular}
\end{table}

Attempts to synthesize the superheavy elements 119 and 120 were performed at different laboratories in the world, i.e, FLNR, GSI, GANIL etc. But no decay chains were observed up to now in experiments. Theoretical predictions with various models were also made for producing the element 119 with a number of systems, e.g., $^{249}$Cf ($^{45}$Sc,xn) $^{294-x}$119, $^{251}$Cf ($^{45}$Sc,xn) $^{296-x}$119, $^{247}$Bk ($^{50}$Ti,xn) $^{297-x}$119, $^{249}$Bk ($^{50}$Ti,xn) $^{299-x}$119, $^{254}$Es($^{48}$Ca,xn)$^{302-x}119$ etc. Some results for the optimal ER cross sections in different reactions leading to the formation of Z=119 are listed in Table I. One can observe that the optimal system of $^{252}$Es ($^{44}$Ca,3n) $^{293}119$ is possible with a larger cross section of 4 pb by the DNS model \cite{Li18}. Further confirmation with different models is expected for checking the reliability of the calculation. On the other hand, the construction of target material $^{252}$Es in experiment is very difficulty. The reaction of $^{45}$Sc+$^{249,251}$Cf is also feasible for synthesizing the new element with the cross section above 0.1 pb. Difference of one order magnitude for producing the element 119 in the reaction $^{50}$Ti+$^{249}$Bk exists in the model predictions, e.g., 0.03 pb in the 4n channel at the excitation energy of 36 MeV by the FBD model \cite{Si46}, 0.57 pb at 41 MeV by the diffusion model with Langevin-type equations \cite{Liu11}, 0.11 pb in Ref. \cite{Wa12} and 0.013 pb in our calculation by the DNS model. The 3n and 4n channels are optimal way to create the new element as shown in Fig. 5. The isotopic dependence of ER cross sections is weak. More large mass asymmetry in the bombarding system is favorable to produce SHN. The synthesis of element 119 in laboratories is possible in the forthcoming experiments with the high-intensity accelerators in the world. The reliable prediction in theories is helpful for experimental managements.

\begin{figure*}
\includegraphics[width=14 cm]{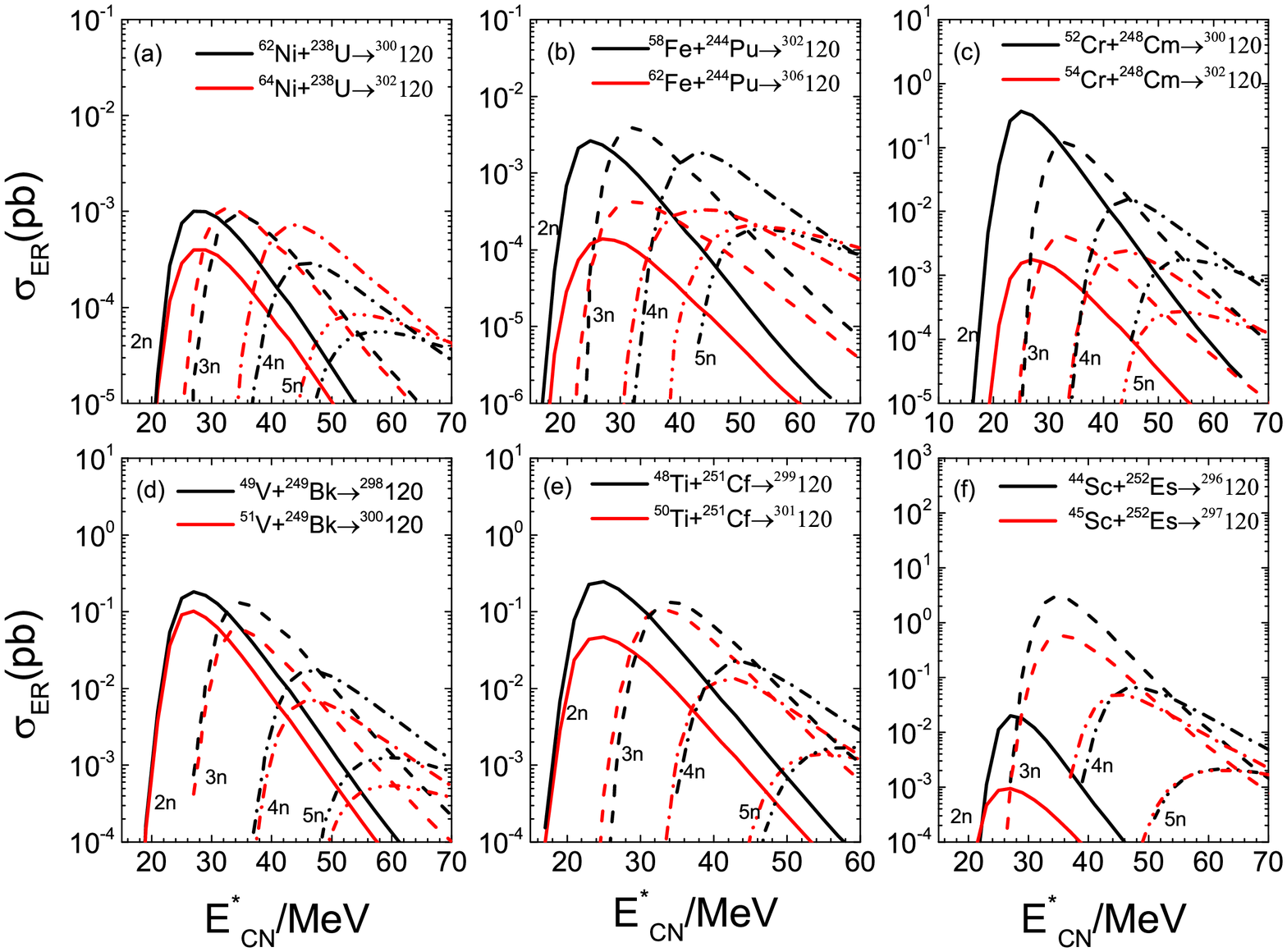}
\caption{ The same as in Fig. 5, but for the production of element Z=120. }
\end{figure*}

\begin{table}[!htb]
	\centering
	\caption{The same as in table I, but for the production of SHE Z=120.}
	\label{tab2}
		\begin{tabular}{ccccccc}
			Reaction systems     &$\sigma_{ER}(pb)$     &$E_{CN}^{\ast}$(MeV)     &References \\ \hline
	       $^{249}$Cf($^{50}$Ti,4n)$^{295}$120 &0.006      &43           &\cite{Si46}   \\
		   $^{251}$Cf($^{50}$Ti,4n)$^{297}120$  &0.003      &42      \\
           $^{248}$Cm($^{54}$Cr,4n)$^{298}120$ &0.001   &35       \\
            \cline{1-4}
            $^{244}$Pu($^{58}$Fe,3n)$^{299}120$  &0.01  &36  &\cite{Si50} \\
            $^{238}$U($^{64}$Ni,3n)$^{299}120$  &0.007  &36     \\
            $^{248}$Cm($^{54}$Cr,3n)$^{299}120$ &0.076  &36   \\
            $^{249}$Cf($^{50}$Ti,3n)$^{296}120$ &0.76  &33         \\
            \cline{1-4}
            $^{249}$Cf($^{50}$Ti,3n)$^{296}120$ &0.1  &29   &\cite{Na51} \\
            $^{248}$Cm($^{54}$Cr,3n)$^{299}120$ &0.055  &30   \\
            \cline{1-4}
            $^{249}$Cf($^{50}$Ti,4n)$^{295}120$ &0.046  &43    &\cite{Za08} \\
            $^{248}$Cm($^{54}$Cr,4n)$^{298}120$ &0.028  &43   \\
            \cline{1-4}
            $^{249}$Cf($^{50}$Ti,3n)$^{296}120$ &0.06  &36   &\cite{Li11} \\
            $^{250}$Cf($^{50}$Ti,3n)$^{297}120$ &0.12  &37   \\
            $^{251}$Cf($^{50}$Ti,4n)$^{297}120$ &0.11  &38    \\
            $^{252}$Cf($^{50}$Ti,4n)$^{298}120$ &0.25  &38  \\
            \cline{1-4}
            $^{251}$Cf($^{50}$Ti,3n)$^{298}120$ &0.25  &36   &\cite{Wa12} \\
            $^{249}$Cf($^{50}$Ti,3n)$^{296}120$ &0.05  &33           \\
            $^{248}$Cm($^{54}$Cr,4n)$^{298}120$  &0.005  &42    \\
            $^{244}$Pu($^{58}$Fe,4n)$^{298}120$  &0.003  &43     \\
            $^{249}$Cf($^{50}$Ti,3n)$^{296}120$  &0.02  &31           \\
            \cline{1-4}
            $^{257}$Fm($^{40}$Ca,3n)$^{294}120$ &1.24  &48   &\cite{Li18}  \\
            $^{248}$Cf($^{46}$Ti,2n)$^{292}120$  &0.17  &34   \\
            $^{249}$Cf($^{46}$Ti,3n)$^{292}120$  &0.24  &39   \\
            $^{250}$Cf($^{46}$Ti,2n)$^{294}120$  &0.13  &36   \\
            $^{251}$Cf($^{46}$Ti,3n)$^{294}120$  &0.37  &39   \\
            \cline{1-4}
            $^{251}$Cf($^{50}$Ti,3n)$^{298}120$ &0.11  &33   &This work \\
            $^{251}$Cf($^{48}$Ti,2n)$^{297}120$ &0.25  &25         \\
            $^{244}$Pu($^{58}$Fe,3n)$^{299}120$  &0.004  &33   \\
            $^{244}$Pu($^{62}$Fe,3n)$^{303}120$  &0.0004  &31  \\
            $^{248}$Cm($^{54}$Cr,3n)$^{299}120$  &0.004  &33  \\
            $^{248}$Cm($^{52}$Cr,2n)$^{300}120$  &0.37  &25    \\
            $^{238}$U($^{64}$Ni,3n)$^{299}120$  &0.001  &31   \\
            $^{238}$U($^{62}$Ni,2n)$^{300}120$  &0.001  &27   \\
            $^{252}$Es($^{44}$Sc,3n)$^{293}120$  &3.18  &35   \\
            $^{252}$Es($^{45}$Sc,3n)$^{293}120$  &0.59  &35   \\
            $^{249}$Bk($^{49}$V,2n)$^{296}120$  &0.18  &27   \\
            $^{249}$Bk($^{51}$V,2n)$^{298}120$  &0.1  &27
 		\end{tabular}
\end{table}

The synthesis of superheavy element 120 is particularly important in understanding the shell structure in the domain of SHNs. The strongly shell effect enhances the fission barrier and $\alpha$ decay half-life, which is also favorable for producing and surviving SHN. Possible experiment is being planned at HIAF with high-intensity beams. Shown in Fig. 6 is the systematic comparison for producing SHN of Z=120 with actinide nuclides based reactions. Different panels represent different reaction systems to create the superheavy element 120, but with different neutron richness. The compound nuclei formed by different combinations are close to the neutron shell closure N=184. The channels of 2-4n are available for the SHN production with the excitation energies around 25 MeV, 30 MeV and 40 MeV, respectively. The maximal ER cross section depends on the isotopic projectiles. It is obvious that the 3n channel of the  $^{44,45}$Sc + $^{252}$Es reaction is favorable for synthesizing the new SHN owing to the large mass asymmetry in the entrance system. The smaller neutron separation energy is available for cooling the compound nucleus formed in the fusion reactions. In Table II, the production cross sections of Z=120 with the optimal channels and possible combinations are compared. Difference of model predictions exists in the calculations, e.g., the production of $^{295}$120 in collisions of $^{50}$Ti on $^{249}$Cf being 0.006 pb and 0.046 pb by the FBD model \cite{Si46} and multidimensional Langevin-type equations \cite{Za08}, respectively. Calculations support the 3n channel in collisions of $^{50}$Ti on californium isotopes is available for synthesizing the element 120 with the cross section above 0.1 pb at the excitation energy around 35 MeV. The 2n channel in the reactions of $^{49,51}$V+$^{249}$Bk is also possible for creating the new element. The position of maximal cross section is mainly determined by the odd-even effect of neutron evaporation and the energy dependence of fusion probability. It should be noticed that the proton shell closure Z=120 was predicted with the relativistic mean-field model by including the isospin dependence of the spin-orbit interaction and the effective mass \cite{Be99}. The shell correction energy calculated by the macroscopic-microscopic model is used in the calculation and the Z=114 proton shell closure is given by the approach \cite{Mo95}. The production cross section of element 120 is to be enhanced with the shell correction by the relativistic mean-field model.

\begin{figure*}
\includegraphics[width=14 cm]{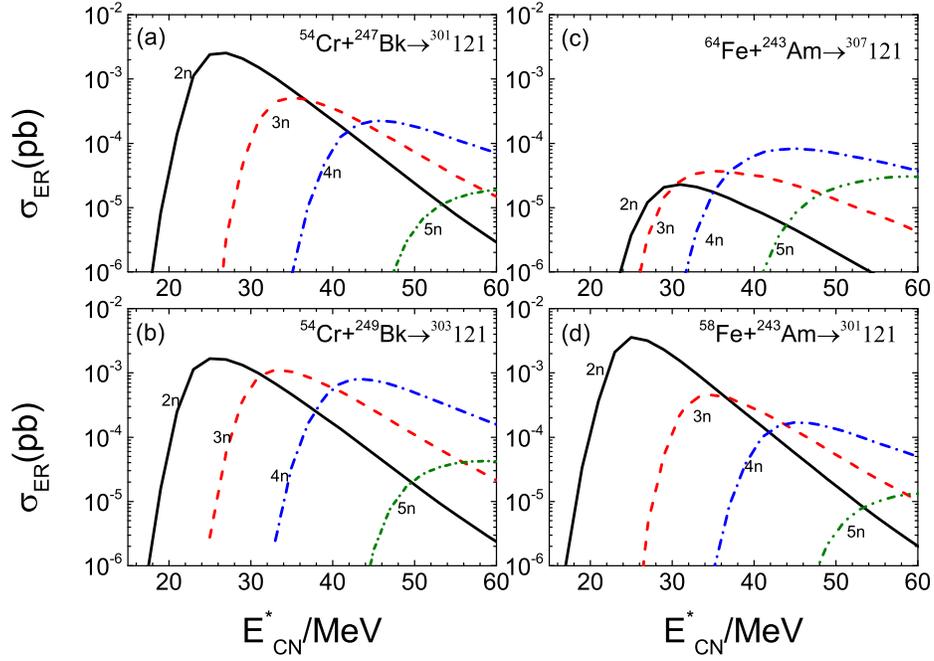}
\caption{ The evaporation residue cross sections with channels of (2-5)n of element Z=121 in collisions by different combinations.}
\end{figure*}

\begin{figure*}
\includegraphics[width=14 cm]{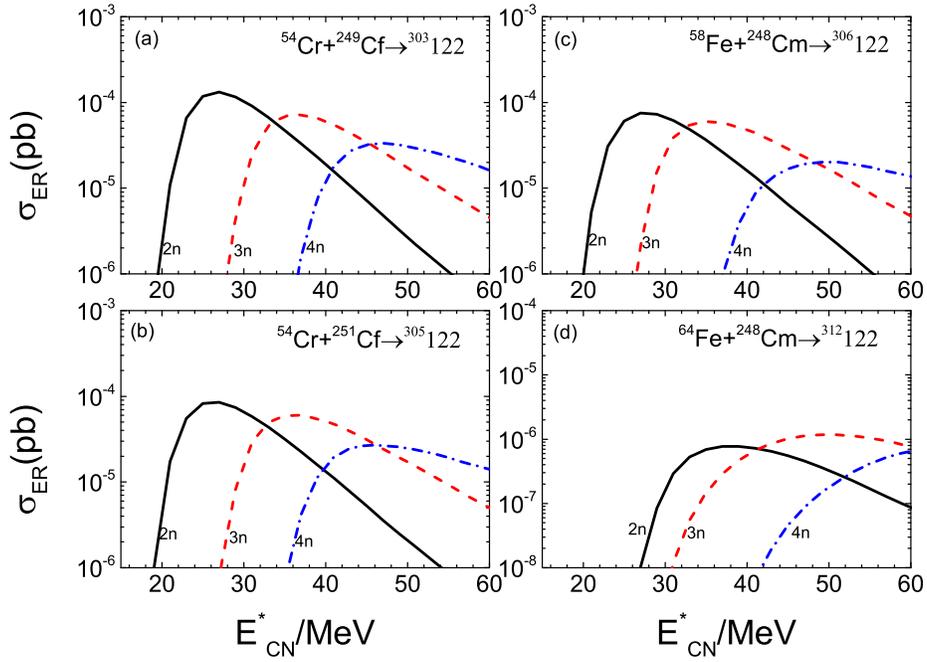}
\caption{ The same as in Fig. 5, but for the production of element Z=122.}
\end{figure*}

As the extension of the model prediction, we analyzed the formation of superheavy elements 121 and 122 in massive fusion reactions. Shown in Fig. 7 is the ER excitation functions in the reactions of $^{54}$Cr+$^{247,249}$Bk and $^{58,64}$Fe+$^{243}$Am. In panels (a) and (b), the stable nuclide $^{54}$Cr and isotopic target nuclei are selected. It can be seen that the 2-4n channels in the $^{54}$Cr+$^{249}$Bk reaction are favorable, but with very low cross section at the level of 1 fb. The influence of stable and radioactive nuclides on the SHN production is compared in panels (c) and (d). The different structure of 2-5 channels is caused from the neutron separation energy in the cascade evaporation, i.e., the smaller separation energy for the compound nucleus $^{301}$121 resulting in the larger 2n channel probability. The production cross section of element 122 is very low in the fusion reactions as shown in Fig. 8. Both the neutron separation energy and fission barrier are sensitive to the survival of SHN. The polar angle distribution, mass, total kinetic energy spectra and excitation energy dependence of fission fragments from the SHNs are useful for extracting the fission barrier and shell evolution. More experiments are expected in the future. The systems of $^{54}$Cr+ $^{249,251}$Cf and $^{58,64}$Fe+$^{248}$Cm are chosen for synthesizing the element 122 as shown in Fig. 8. The cross section below 0.1 fb is already out the limit of experimental measurement. The new reaction mechanism is expected for creating the neutron-rich SHN and new element, i.e., the multinucleon transfer reaction, incomplete fusion with radioactive nuclide etc.

\section{Conclusions}

Within the framework of the DNS model, the SHN formation in the fusion-evaporation reactions is thoroughly investigated. The stochastic collision orientations in the nucleon transfer process are implemented into the model via the Monte Carlo approach. The calculated results are consistent with the experimental data from Dubna. The maximal cross sections of evaporation residues appear in the (2-5)$n$ evaporation channels. The yields in the 1$pxn$, 1$\alpha xn$ and 1$p$1$\alpha xn$ evaporation channels are much lower than the pure neutron evaporation. The reactions of $^{249}$Cf ($^{45}$Sc,xn)$^{294-x}$119, $^{251}$Cf($^{45}$Sc,xn)$^{296-x}$119, $^{247}$Bk($^{50}$Ti,xn) $^{297-x}$119 and $^{249}$Bk($^{50}$Ti,xn)$^{299-x}$119 are investigated for synthesizing the new element 119. It is concluded that the maximum cross sections are close to 1 pb in the 3$n$ evaporation channel for the systems and weakly depend on the isotopic target nucleus. The synthesis of the element $Z=120$ is investigated by a series of isotopic projectile nuclei bombarding on actinide targets. The optimal combination is the reaction of $^{44}$Sc+$^{252}$Es in the 3$n$ channel with the cross section of 3 pb. The production of superheavy elements 121 and 122 is obtained at the level of below 1 fb in the massive fusion reactions. New reaction mechanism is still needed to be explored for the new element production. Therefore, the synthesis of new SHN in experiments provides a good theoretical basis for the selection of collision combination.

\section{Acknowledgements}

This work was supported by the National Natural Science Foundation of China (Projects No. 12175072 and No. 11722546) and the Talent Program of South China University of Technology.

\end{document}